\documentclass[twocolumn,showpacs,prl]{revtex4-1}
\usepackage{graphicx}
\usepackage{dcolumn}
\usepackage{bm}
\usepackage{leftidx}

\begin{document}
\title{Quantum scattering model of energy transfer in photosynthetic complexes}
\author{Bao-quan  Ai$^1$}
\author{Shi-Liang Zhu$^{1,2}$}\email{shilzhu@yahoo.com.cn}
\affiliation{$^1$Laboratory of Quantum Information Technology and
SPTE,South China Normal University, Guangzhou, China\\
$^2$ Center for Quantum Information, IIIS, Tsinghua University}

\begin{abstract}
\indent We develop a quantum scattering model to describe the
exciton transport through the Fenna-Matthews-Olson(FMO) complex.
It is found that the exciton transport involved the optimal
quantum coherence is more efficient than that involved classical
behavior alone. Furthermore, we also find that the quantum
resonance condition is easier to be fulfilled in multiple pathways
than that in one pathway. We  then definitely demonstrate that the
optimal distribution of the pigments, the multitude of energy
delivery pathways and the quantum effects, are combined together
to contribute to the perfect energy transport in the FMO complex.
 \end{abstract}

 \maketitle

Photosynthesis is a fundamental biological process that harvests
solar energy to power life on Earth. It starts with the absorption
of a photon of sunlight by the antenna molecules, followed by
transfer of the energy to the reaction centers. Remarkably, the
energy transfer from the antennas to the reaction center is almost
perfectly efficient\cite{a1}. Then an intriguing question arises:
what makes the energy transfer so efficient?   Up to date, there
are two acceptable contributing factors\cite{a2}: The first one is
optimal space distribution of the pigments in light-harvesting
complexes\cite{a3}, where the optimal space is just closely enough
to enable fast energy transfer, but far enough apart to prevent
the molecular orbital of the pigments from overlapping. The second
factor is the supramolecular organization of photosynthetic
apparatus, which allows the multiple energy delivery pathways to
connect to the reaction centre\cite{a4}.

Recently, the experimental and theoretical studies show that there
exists long-lived quantum coherence in this energy transfer in
several photosynthetic complexes \cite{a5,a6,a7,a8,a9,a10}, which
suggests that quantum coherence could be a third factor in
optimizing energy transfer efficiency.  However, to what extent
quantum effects contribute to the efficiency of energy transfer is
largely unknown. Therefore, it would be significant important to
establish quantitatively whether or not photosynthetic light
harvesting that involves quantum coherence is truly  more
efficient than it would be using classical mechanisms
alone\cite{a2}.

In this paper, we develop a quantum scattering model to describe
the exciton (the energy carrier) transport in the FMO complex. We
explore respectively the quantum and classical transports of the
exciton in one, two and multiple pathways and show that the
quantum transport under the resonant condition can lead to perfect
efficiency. Since the resonance is due to the quantum coherence
and it is impossible in classical physics, we thus demonstrate
definitely that the exciton transport involved optimal quantum
coherence is more efficient than that for the classical case. We
show that the resonant transport with perfect efficiency occurs
when the quantum phase of the exciton accumulated in the pathways
are fulfilled with the resonant condition. Since the phase of the
exciton is determined by the space distribution of the pigments,
the optimal distribution of the pigments is then indeed
significant in the near-unity transfer efficiency. Although the
perfect efficiency is also possible for one pathway alone, the
multiple pathways are essential because the resonant region in
multiple pathways is much larger than that in one pathway.  The
large resonant region is the mechanism that the excitons with a
broad range of the spectrum ( electron absorbed different colors
of light) can transport  with perfect efficiency. Therefore, the
three mentioned contributing factors, the optimal distribution of
the pigments, the multitude of energy delivery pathways and the
quantum effects, are combined together in our proposed model to
contribute the perfect energy transport in FMO.

\begin{figure}[htbp] \includegraphics[height=4cm]{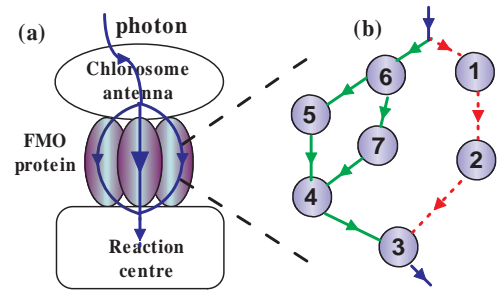}
\caption{(Color online) The schematic representation of the
light-harvesting apparatus of green sulphur bacteria. (a) The
light is absorbed by the chlorosome antenna and then transported
to the reaction center through the FMO
complex. (b) Each monomer has seven pigments labelled by
$\textcircled1-\textcircled7$. Two main energy transfer pathways
$L_A$ and $L_B$  are indicated by the red and green arrows
\cite{a11}, $L_{A}$:
$\textcircled1$$\rightarrow$$\textcircled2$$\rightarrow$$\textcircled3$;
$L_{B}$: a ring structure contains two paths
$\textcircled6$$\rightarrow$$\textcircled5$$\rightarrow$$\textcircled4$$\rightarrow$$\textcircled3$
and
$\textcircled6$$\rightarrow$$\textcircled7$$\rightarrow$$\textcircled4$$\rightarrow$$\textcircled3$.
}
\end{figure}

\emph{Quantum scattering model of the FMO complex.}
The architecture of antenna
light-harvesting complexes varies widely among photosynthetic
organisms. A well-studied example is the water-soluble FMO complex
of green sulfur bacteria.  As shown in Fig. 1, the FMO complex
comprises three identical monomers that each contains seven
pigments. The seven pigments (Fig. 1(b)) can be combined into two
major energy transport pathways \cite{a11}. A light pulse absorbed
by the chlorosome antenna could excite an electron into a state
called an exciton which is a quasi-particle and can be considered
as the energy carrier in light-harvesting complexes. The exciton
then transports to the reaction centre through FMO with nearly
perfect efficiency.

\indent Exciton transport is accomplished in FMO through
tunneling, where the exciton jumps from one pigment to another. In
the previous theories\cite{a9,a10,a12}, each pigment is usually
modeled by a two-level molecule and then the FMO complex are
described by some coupled two-level molecules interacting with
their environmental phonons. In this paper we proposed a different
model, the scattering model\cite{b1,b2} (see a brief review in
Supplementary material), to explain transmission efficiency of
FMO. There are two basic elements in the scattering model, the
scatterer and the ideal channel; they are connected with certain
orders to form a quantum network. We denote the transfer matrix
for the $n$th scatterer as $M_{n}$ and the transfer matrix for the
ideal channel with the length $l$ as $U_l$. They can be written as
\begin{equation}\label{Mn}
     M_{n}=\left(
  \begin{array}{ccc}
    \frac{1}{t_{n}^{*}} & -\frac{r_{n}^{*}}{t_{n}^{*}} \\
    -\frac{r_{n}}{t_{n}} & \frac{1}{t_{n}}  \\
  \end{array}
\right),\ \ \
U_{l}=\left(\begin{array}{cc} e^{-ikl}& 0\\ 0 & e^{ikl}\\
\end{array}\right),
\end{equation}
where $t_{n}$=$T^{\frac{1}{2}}_{n}e^{i\varphi_{n}}$,
$r_{n}$=$-iR_{n}^{\frac{1}{2}}e^{i\varphi_{n}}$, $T_{n}$ is the
transmission probability of the $n$th scatterer, $R_{n}$=$1$-$T_{n}$
is its reflection probability, $\varphi_{n}$ is the phase change
in the transmitted wave and $k$ is the wave number of the exciton.

\begin{figure}[htbp] \includegraphics[width=6cm]{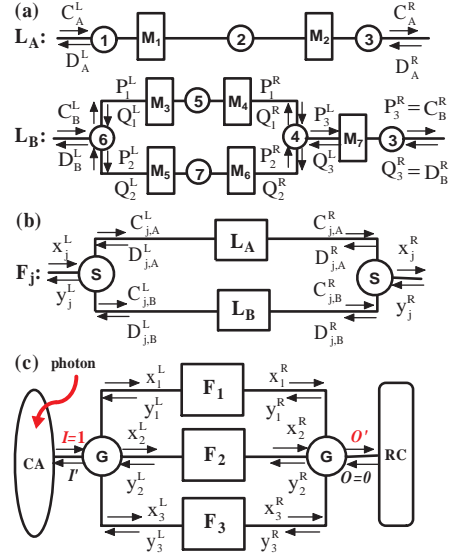}
\caption{The quantum scattering model we proposed to describe the
excitation transmission of FMO. (a)  The scattering matrix
representation of two major pathways $L_A$ and $L_B$. (b) Each
monomer is represented by the symbol $F_j$ made by a ring with two
branches $L_A$ and $L_B$ and connected to two leads. (c) The
scattering matrix representation of the whole FMO complex. The
symbols near the arrows denote the corresponding transmitted and
reflected amplitudes. }
\end{figure}

The transport of FMO  can be well explored with the help of the
transfer matrices in Eq. (1) as well as the junctions $S$ and $G$
to be defined below. Each pigment can be thought of as a potential
well and there exists a potential barrier between two nearest
neighbor pigments along the pathways. The potential barrier can be
well described by the transfer matrix of the scatterer defined in
Eq.(\ref{Mn}).
The wave function of the exciton in
pigment $j$ could be written as $\psi_j (x)$=$C_j e^{ikx}$+$D_j
e^{-ikx}$, where $C_j$ and $D_j$ are the amplitudes of the wave
function. The amplitudes $C_j$ and $D_j$ are related to its
nearest neighbor pigment through the transfer matrix of the
scatterer between the two pigments. In addition, the transfer
matrix of the pigment is the one as the ideal channel defined in
Eq.(\ref{Mn}). Therefore, we can describe the exciton transmission
through the whole FMO complex with the following three steps: (i)
the two major energy transport pathways in each monomer is
described by two different scattering pathways $L_{A}$ and
$L_{B}$, as shown in Fig. 2(a). (ii) Each monomer is represented
by a ring with two branches $L_A$ and $L_B$ and connected to two
leads through two junctions $S$. The transfer matrix of the
monomer $j$ ($j$=$1,2,3$) is denoted as $F_j$, as show in Fig.
2(b). (iii) The transport of the whole FMO can be described in
Fig. 2(c), where three monomers represented by $F_j$ are connected
to the antennas and reaction center through two junctions $G$. The
exciton created at the antennas injects into FMO with $I=1$
through the G junction connected with the antennas. For
simplicity, we assume that there is no reflection from the
reaction center (the amplitude $O$ in Fig. 2(c) vanishes); In this
case, the transmission coefficient of the whole FMO complex is
then determined by the probability $|O'^2|$ indicated in Fig.
2(c).

\emph{Transfer matrices of the FMO complex.} We now turn to study
the transfer matrices of the basic elements $L_A$, $L_B$, $F_j$
and the whole FMO complex. The pathway $L_{A}$ is represented by
two scatterers connected one ideal channel
 and its transfer matrix is then given by
\begin{equation}\label{}
L_{A}=M_{2}U_{l}M_{1},\
\end{equation}
where $l$ is the length of the 2nd pigment. The pathway $L_{B}$ is
composed of three branches connected by two $S$ junctions
($\textcircled6$ and $\textcircled4$). For each branch, we have
\begin{equation}
\label{Lj}
 \left(
  \begin{array}{cc}
    P^{R}_m \\
    Q^{R}_m \\
  \end{array}
\right)= \Gamma_{m} \left(
  \begin{array}{cc}
    P^{L}_m\\
    Q^{L}_m \\
  \end{array}
\right), (m=1,2,3)\end{equation} where
$\Gamma_1$=$U_{l_{1}}M_{4}U_{l_{1}}M_{3}U_{l_{1}}$,
$\Gamma_2$=$U_{l_{2}}M_{6}U_{l_{2}}M_{5}U_{l_{2}}$ and
$\Gamma_3$=$M_{7}U_{l_{3}}$. Here $l_{m}$ is the distance between
the scatterers or the distance between the scatterers and the
junctions for the $m$th branch. The amplitudes of the three
outgoing waves of the junction $S$ are related by a $S$ matrix to
the amplitudes of the three incoming waves, i.e.,
\begin{equation}\label{}
\left(
  \begin{array}{ccc}
    D_{B}^{L} \\
    P_{1}^{L} \\
    P_{2}^{L}\\
  \end{array}
\right)=S \left(
  \begin{array}{ccc}
     C_{B}^{L}\\
    Q_{1}^L \\
    Q_{2}^L\\
  \end{array} \right), \ \ \
  \left(
  \begin{array}{ccc}
    P_{3}^{L} \\
    Q_{1}^{R} \\
    Q_{2}^{R}\\
  \end{array}
\right)=S \left(
  \begin{array}{ccc}
     Q_{3}^{L}\\
    P_{1}^{R} \\
    P_{2}^{R}\\
  \end{array}
\right),
\end{equation}
where $S$ is a $3\times 3$ unitary matrix and its detailed form is
given by Eq. (14) in Supplementary material.

\indent From Eqs. (1-4), the amplitudes $C^L_\alpha$,$D^L_\alpha$
$(\alpha$=$A,B)$ of the wave function on the left of the pathway and
the amplitudes $C^R_\alpha$,$D^R_\alpha$ on the right of the
pathway can be related by the transfer matrixes $L_{\alpha}$,
that is,
\begin{equation}
\label{Lj}
 \left(
  \begin{array}{cc}
   C^{R}_\alpha \\
    D^{R}_\alpha \\
  \end{array}
\right)= L_{\alpha} \left(
  \begin{array}{cc}
    C^{L}_\alpha\\
    D^{L}_\alpha \\
  \end{array}
\right).\end{equation}
 Therefore, instead of an array of the
scatterers, each pathway may be replaced by an effective transfer
matrix.

\indent We now study the exciton transport in one monomer of the
trimer. The schematic representation of the monomer is shown in
Fig. 2(b). Two pathways $L_A$ and $L_B$ form a ring and connected
to two leads through two $S$ junctions.
 With the help of the effective transfer matrix for the pathway $\alpha$ of the monomer $j$, we can obtain
 \begin{equation}
\label{Lj}
 \left(
  \begin{array}{cc}
    C^{R}_{j,\alpha} \\
     D^{R}_{j,\alpha} \\
  \end{array}
\right)= L_{j,\alpha} \left(
  \begin{array}{cc}
    C^{L}_{j,\alpha}\\
    D^{L}_{j,\alpha} \\
  \end{array}
\right),\ \ (j=1,2,3)\end{equation}
 where $L_{j,\alpha}$=$U_{l_{j,\alpha}}M_{j,\alpha}U_{l_{j,\alpha}}$, and $l_{j,\alpha}$ is the distance
 between the scatterers and $S$ junctions for the pathway $\alpha$ of the monomer $j$.

The amplitudes of the three outgoing waves are related to the
amplitudes of the three incoming waves through $S$ matrix given by
\begin{equation}\label{}
\left(
  \begin{array}{ccc}
    y_{j}^{L} \\
    C_{j,A}^{L} \\
    C_{j,B}^{L}\\
  \end{array}
\right)=S \left(
  \begin{array}{ccc}
     x_{j}^{L}\\
    D_{j,A}^L \\
    D_{j,B}^L\\
  \end{array} \right), \left(
  \begin{array}{ccc}
    x_{j}^{R} \\
    D_{j,A}^{R} \\
    D_{j,B}^{R}\\
  \end{array}
\right)=S \left(
  \begin{array}{ccc}
     y_{j}^{R}\\
     C_{j,A}^{L} \\
     C_{j,B}^{L}\\
  \end{array}
\right).
\end{equation}
From Eqs. (6-7), the amplitudes of the wave
function on the left of the monomer $j$ and the amplitudes on the
right are related by the matrix $F_j$ given by
\begin{equation}
\label{Fj} \left(
  \begin{array}{cc} x^R_j\\
    y^{R}_j \\
  \end{array}
\right)= F_{j} \left(  \begin{array}{cc}
    x^{L}_j\\
    y^{L}_j \\
  \end{array}
\right).\end{equation}

Similar to the transmission of one monomer, we can describe the
exciton transport through FMO by two $G$ junctions as shown in
Fig. 2 (c). The amplitudes of the four outgoing waves of the
junction $G$ are related by a $G$ matrix to the amplitudes of the
four incoming waves
\begin{equation}\label{M_FMO}
\left(
  \begin{array}{c}
    I^{'} \\
    x_{1}^{L} \\
    x_{2}^{L}\\
    x_{3}^{L} \\
  \end{array}
\right)=G \left(
  \begin{array}{c}
     I \\
    y^{L}_1 \\
    y^{L}_2\\
    y^{L}_3\\
  \end{array}
\right),\ \ \  \left(
  \begin{array}{c}
    O^{'} \\
    y_{1}^{R} \\
    y_{2}^{R}\\
    y_{3}^{R} \\
  \end{array}
\right)=G \left(
  \begin{array}{c}
     O \\
    x^{R}_1 \\
    x^{R}_2\\
    x^{R}_3\\
  \end{array}
\right),
\end{equation}
where $G$ is a $4 \times 4$  unitary matrix given by Eq. (20) in
Supplementary material. By solving Eq.(\ref{M_FMO}) with $I$=$1$ and
$O$=$0$, one can obtain the transmission coefficient
$T_{FMO}$=$|O'^2|$ of the FMO complex.

To understand the features of the transmission, we compare the
transmission coefficients for different number of pathways. For
simplicity, we assume that all scatterers have the same
transmission probability $T_{n}$=$T$ and the phase change in
transmitted wave $\varphi_{n}$=$0$. Furthermore,  we compare the
transmission difference between the classical and quantum cases.
As for the classical transmission, it means that the exciton is a
classical particle without wave-like phase and thus the particle
probability is determined by the addition of the probability from
different pathways and not the amplitudes of the wave function; In
addition, the scatterer is determined uniquely by the transmission probability $T$.

\emph{Transmission of the pathways $L_A$ and $L_B$.} Based on Eqs.
(1,2,5), we can obtain the total transmission probability
$T_{A}$=$|C_{A}^{R}|^{2}$ for pathway $L_{A}$ under the conditions
$C_{A}^{L}$=$1$ and $D_{A}^{R}$=$0$,
\begin{equation}\label{}
\small T_{A}={T^{2}}/{[T^{2}+4(1-T)\cos^{2}\phi_{A}]},
\end{equation}
where $\phi_{A}$=$kl$. It is notable that the resonant tunneling
$T_{A}$=$1$  appears under the condition $\phi_{A}$=$(n+1/2)\pi$
with $n$ being an integer. However, as for the classical case,
after considering the multi-reflection between the scatterers
$M_1$ and $M_2$, we may obtain the total transmission probability
$T^{C}_A$=$\frac{T}{2-T}$.

\indent Figure 3(a) shows the total transmission probability as a
function of the path phase for both quantum and classical case at
strong scattering limit $T$=$0.01$. For classical case,
$T^{C}_A$=$5\times10^{-3}$, it is impossible for the particle to
pass through the system at strong scattering limit. However, for
quantum case, there exist two resonant peaks at which the total
transmission probability takes its maximal value $T_{A}$=$1$.
Therefore, the transport involved optimal quantum coherence is
more efficient than that involved the classical behaviors alone.

For pathway $L_{B}$, we set $C_{B}^{L}$=$1$, $D_{B}^{R}$=$0$, $kl_{1}$=$kl_{2}$=$\phi_{B}$ and $kl_{3}$=$\phi_{B}^{'}$. From Eqs. (2-5),
 the total transmission probability $T_{B}$=$|C_{B}^{R}|^{2}$ is
 given by
\begin{equation}\label{}
\small
   T_{B}=\frac{T^{3}}{T^2\sin^2\gamma+[(1+R)\cos\gamma + 2R\cos\phi_{B}^{'}]^{2}},
\end{equation}
where $\gamma$=$2\phi_{B}$+$\phi_{B}^{'}$. The resonant tunneling
$T_{B}$=$1$ will happen under the conditions $\gamma$=$\pi$ and $\phi_{B}$=$\frac{1}{2}\arcsin[(T^{\frac{3}{2}}$+$T$-$2)/2R]$.

\begin{figure}[htbp]
\label{Fig1}\includegraphics[height=3.1cm]{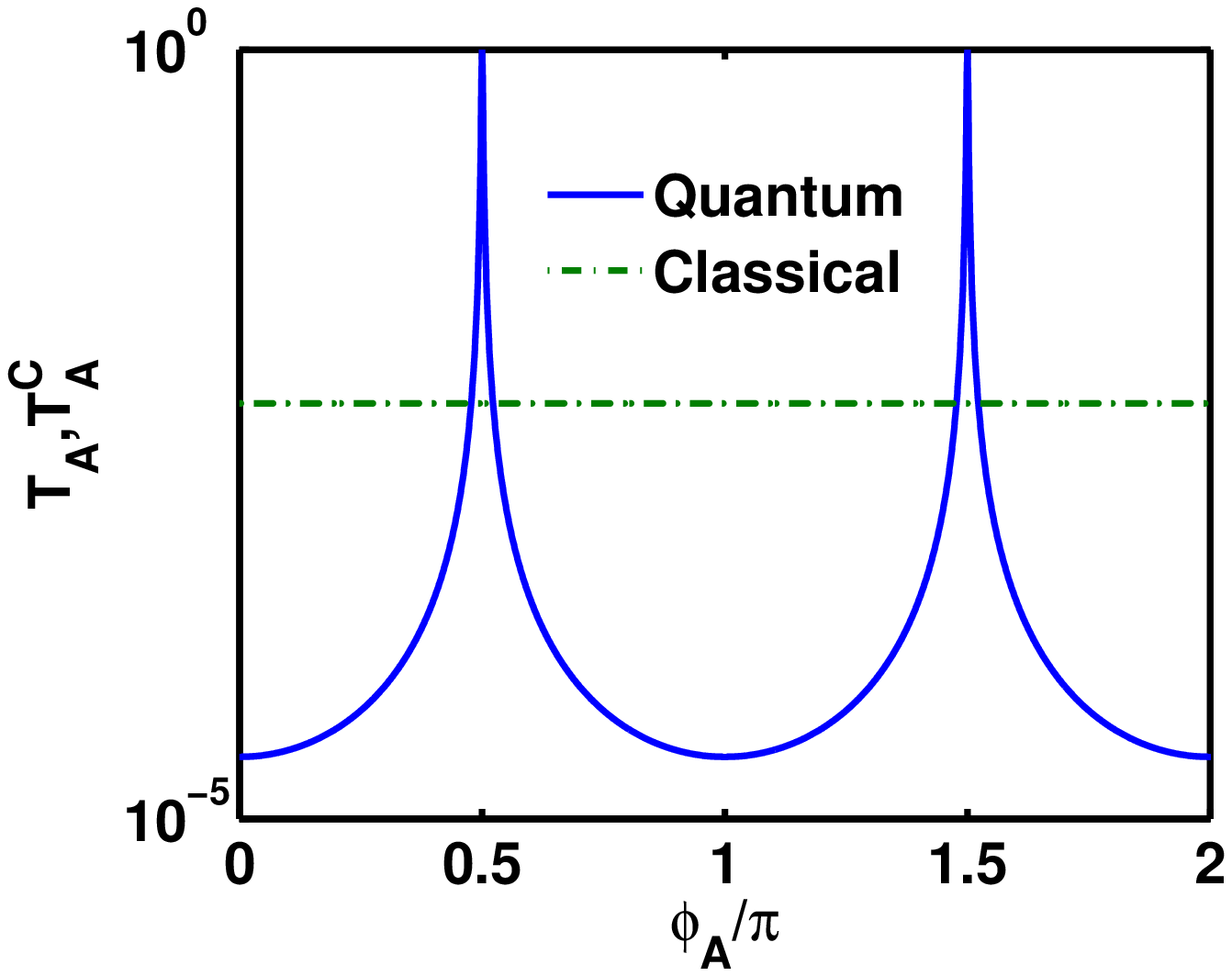}
\includegraphics[height=3.1cm]{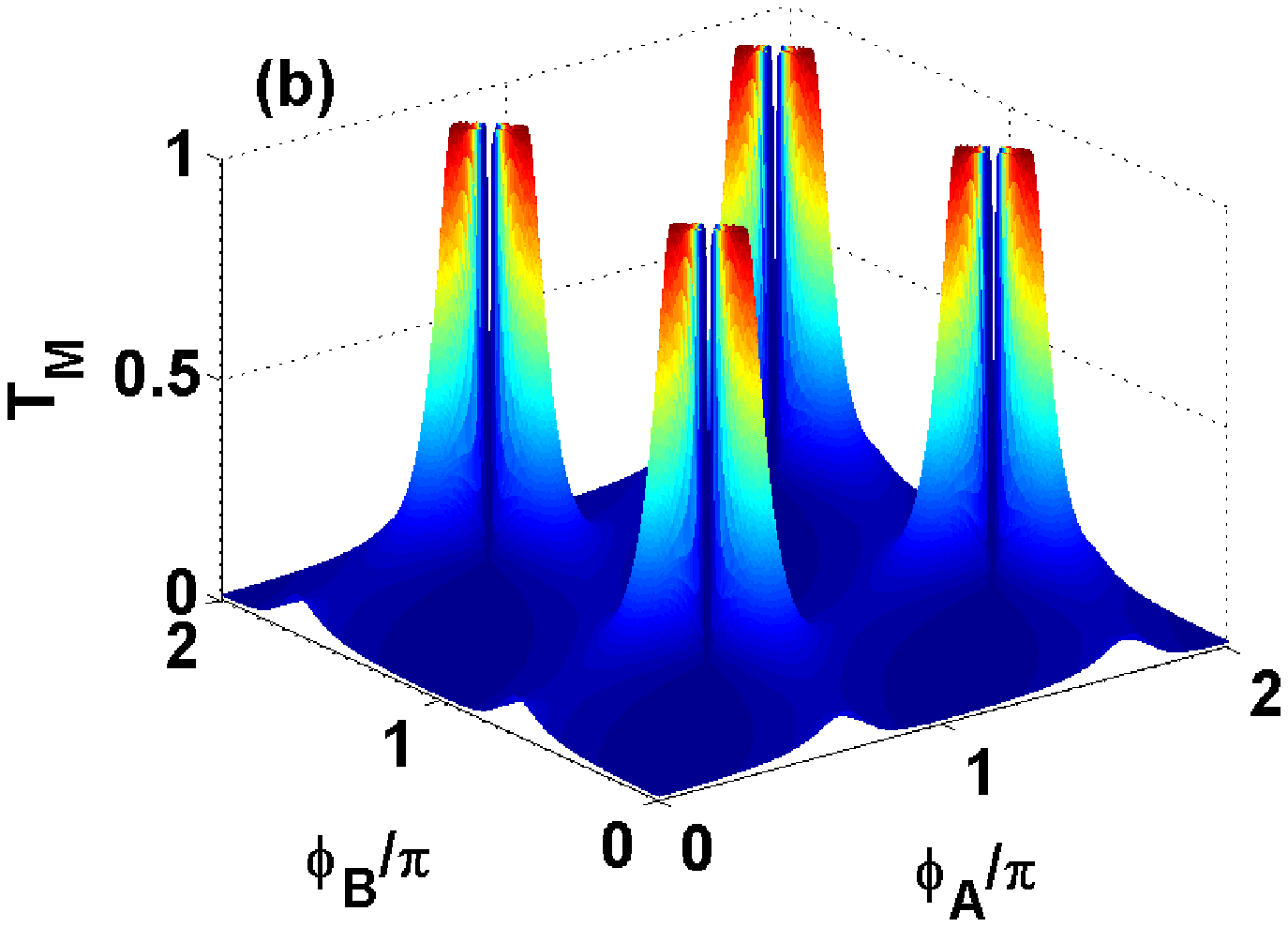}
\includegraphics[height=3.1cm]{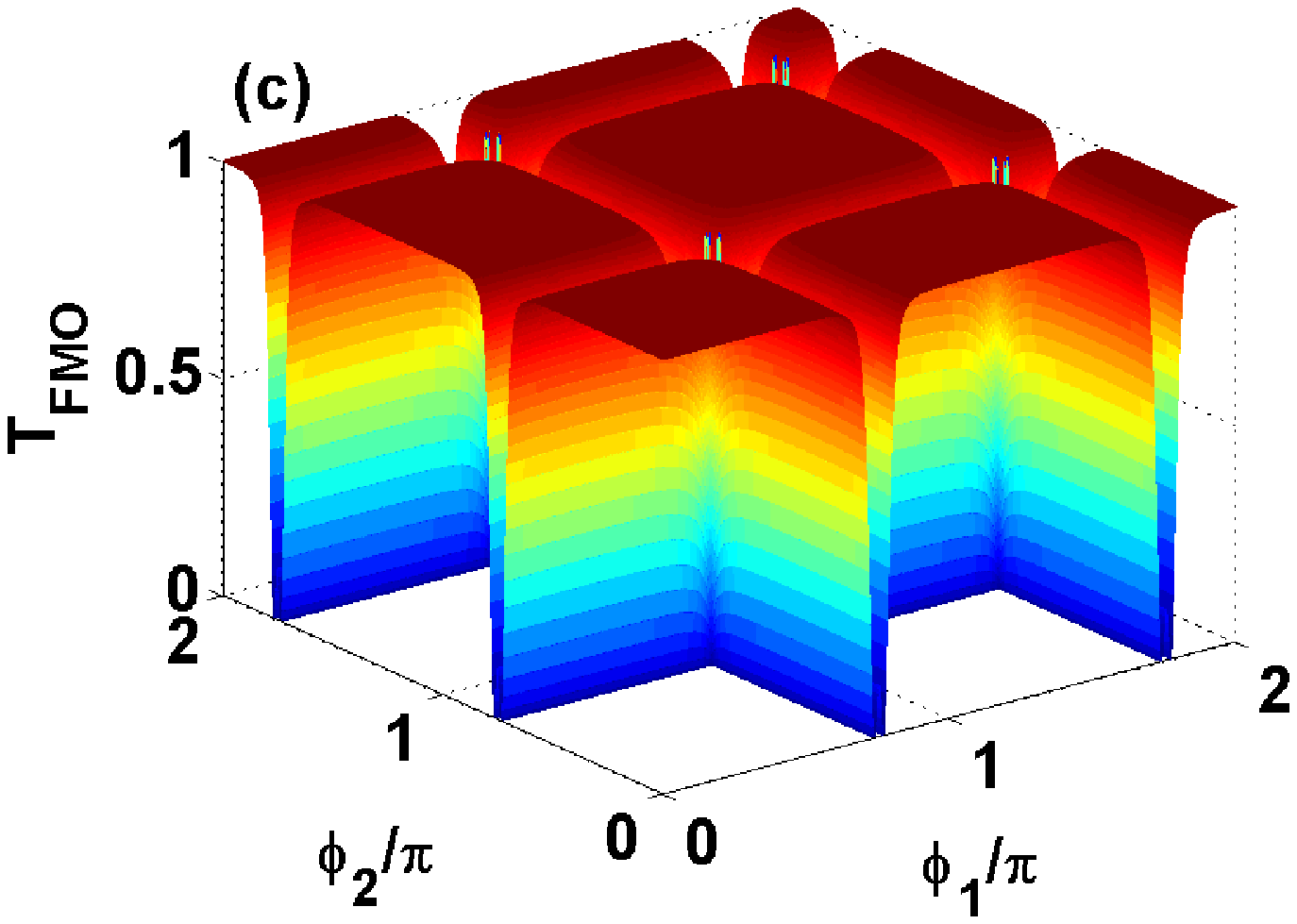}
\includegraphics[height=3.1cm]{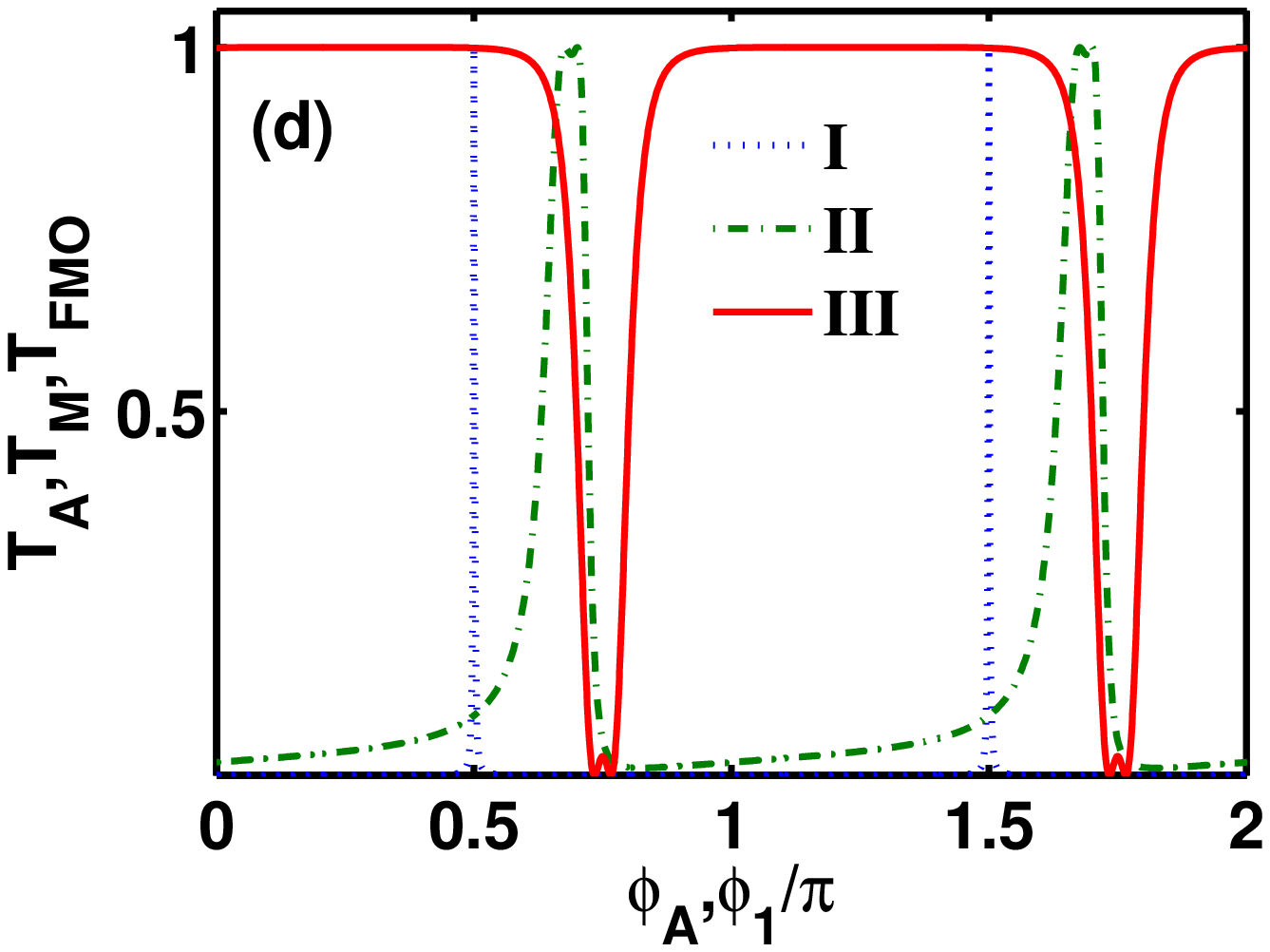}
\caption{(Color online) The transmission probability as a function
of the path phases at $T$=$0.01$. (a) Single pathway $L_{A}$ for both quantum
and classical cases.  (b) A monomer with double pathways.
 (c) The whole FMO with multiple pathways at $\phi_{3}$=$\frac{1}{2}\arcsin(\sqrt{0.12T}$-$\sqrt{R})$.
 (d) The resonant structures: (I) one pathway $L_{A}$, (II) two pathways at $\phi_{B}$=$\frac{1}{2}\arcsin(\sqrt{T}$-$\sqrt{R})$,
 (III) multiple pathways at $\phi_{1}$+$\phi_{2}$=$\pi/2$ and $\phi_{3}$=$\frac{1}{2}\arcsin(\sqrt{0.12T}$-$\sqrt{R})$.}
\end{figure}

\indent The path phases $\phi_{A}$, $\phi_{B}$ and $\phi_{B}^{'}$
denote the distance between the neighbor scatterers or the
distance between the scatterers and junctions, which are
determined actually by the space distribution of the seven
pigments. When the pigments are optimally spaced such that
$\phi_{A}$=$\frac{1}{2}\pi$ for $L_{A}$ and $\gamma$=$\pi$,
$\phi_{B}$=$\frac{1}{2}\arcsin[(T^{\frac{3}{2}}$+$T$-$2)/2R]$ for
$L_{B}$, the quantum resonant tunneling will happen and the
exciton can pass through the barriers with perfect efficiency.
Therefore, we have theoretically demonstrated that optimally
spaced pigments in light-harvesting complexes can facilitate the
energy transfer.

\emph{Transmission of one monomer.} From Eqs. (6-8), we can obtain
the total transmission probability of one monomer
$T_{M}$=$|x_{j}^{R}|^{2}$ at $x_{j}^{L}$=$1$, $y_{j}^{R}$=$0$,
$kl_{j,A}$=$\phi_{A}$ and $kl_{j,B}$=$\phi_{B}$, which is given by
\begin{equation}\label{}
\small
    T_{M}={H^{2}}/({P^{2}+Q^{2}}),
\end{equation}
where
$H$=$\sqrt{T}(2\sqrt{R}$+$\sin2\phi_{A}$+$\sin2\phi_{B})$,$P$=$2\sin^{2}(\phi_{A}$+$\phi_{B})\linebreak[1]$-$\sin^{2}(\phi_{A}$-$\phi_{B})
$+$\sqrt{R}(\sin2\phi_{A}$+$\sin2\phi_{B})$,$Q$=$\sin2(\phi_{A}$+$\phi_{B})$+\linebreak[1]$\sqrt{R}(\cos2\phi_{A}$+$\cos2\phi_{B})$.
We can find that the resonant tunneling $T_{M}$=$1$ under the
conditions $\phi_{A}$+$\phi_{B}$=$\frac{\pi}{2}$ and
$\phi_{A}$=$\frac{1}{2}\arcsin(\sqrt T$-$\sqrt R)$. In contrast,
the classical transmission probability $T^{C}_M$ is less than its
upper bound $\frac{3T}{1+2T}$.

\indent Figure 3(b) shows the total transmission probability of
one monomer versus the path phases at $T$=$0.01$. For classical
case, $T_{C}$$<$$0.029$, the particle passes through the monomer with
very small probability. However, for quantum case, there are eight
resonant regions where the particle can pass through the monomer
with perfect efficiency.
 Compared with  single pathway, the resonant regions for two pathways become larger and then it is much easier for the
 exciton absorbed different colors of light
 to pass through the monomer.


\emph{ Transmission of the whole FMO complex.} From Eqs. (6-9), we
can obtain the total transmission probability
$T_{FMO}$=$|O^{'}|^{2}$ of the whole FMO complex. For simplicity,
we assume $kl_{j,A}$+$kl_{j,B}$=$\frac{\pi}{2}$ and the phases
accumulated between $G$ and $S$ are an integer of $2\pi$. Then we
have
\begin{equation}\label{}
\small
    T_{FMO}={H^{2}}/{Q^{2}},
\end{equation}
where $H$=$6\sqrt{T}q_1 q_2
q_3(3R+2p_{1}\sqrt{R}+\frac{1}{2}p_{1}^{2}-\frac{1}{2}p_{2})$,
$Q=-6p_{1}R^{\frac{5}{2}}$
$-(11p_{1}^{2}-6p_{2}+9)R^{2}-2[30p_{0}-7p_{3}+p_{1}(6+7p_{2})]R^{\frac{3}{2}}-(7p_{1}^{2}+34p_{0}p_{1}+4p_{2}^{2}-3p_{2}-4p_{4})R+(9p_{0}+2p_{1})(p_{2}-p_{1}^{2})R^{\frac{1}{2}}
-9p_{0}^{2}-2p_{1}p_{0}-\frac{1}{2}p_{2}^{2}+\frac{1}{2}p_{4}$.
Here $p_{0}$=$\sin2\phi_{1}\sin2\phi_{2}\sin2\phi_{3}$,
 $p_{n}$=$\sin^{n}2\phi_{1}$+$\sin^{n}2\phi_{2}$+$\sin^{n}2\phi_{3}$ ($n$=$1,2,3,4$), and $q_j$=$\sin2\phi_{j}$+$\sqrt{R}$ with $\phi_{j}$=$kl_{j,A}$.
 It is found that the exciton can resonantly pass through FMO  when $\phi_{1}$=$\phi_{2}$=$\phi_{3}$=$\frac{1}{2}\arcsin(\sqrt{T}$-$\sqrt{R})$.
  In contrast, the total classical transmission probability $T^{C}_{FMO}$ is less than its upper bound $\frac{9T}{1+8T}$.
  Figure 3(c) shows the quantum transmission of the whole FMO complex for fixing
 $\phi_{3}$=$\frac{1}{2}\arcsin(\sqrt{0.12T}$-$\sqrt{R})$ and $T$=$0.01$.
 Compared with the transmission of double pathways shown in Fig. 3(b), the resonant region of the whole FMO becomes larger.
 However, the transmission $T^{C}_{FMO}$$<$$0.083$ for classical
 case, which is pretty small.

Before ending, we make several remarks on the main results  that
can be obtained from the scattering model: (i) The mechanism
underlying the perfect energy efficiency is the resonant
transmission induced by the quantum coherent effects, which is
impossible for the classical mechanism. (ii) Multi-pathway plays
significant role in achieving remarkable efficiency of the energy
transfer. To understand this point, we plot in Fig. 3(d) the
transmission coefficients for one, two, and multiple pathways. The
resonant regions for FMO with multiple pathways almost occupy all
parameter space which ensures that the exciton can pass through
FMO with perfect efficiency. The large resonant region means that
the exciton absorbed different colors of light can still pass
through FMO  with high efficiency. (iii) The resonant conditions
are determined by the space distribution of the pigments. When the
pigments are optimally spaced such that the quantum resonant
tunneling will happen, the exciton can pass through FMO with
perfect efficiency. Therefore, our results prove the
 conjecture that quantum coherence might  'wire' together the final energy acceptors in antennas, thus
 compensating for the relatively weak coupling between the pigments in FMO \cite{a2,a8}. In addition, FMO with multiple
pathways has fault-tolerant function. When one pathway or one
subunit is blocked, the exciton can still pass through other
pathways.

In summary, we have proposed a scattering model to compare the
quantum and classical energy transfer in FMO. We found that, the
aforementioned three factors, the optimal space distribution of
the pigments, multiple pathways, and quantum effects are combined
together to ensure high efficiency of the energy transfer in FMO.

\indent This work was supported by the NNSFC (Nos.11175067, and 11125417), the NBRPC (No.2011CBA00302), the SKPBRC
(No.2011CB922104), and the NSF of Guangdong (No.S2011010003323).

\end{document}